\documentclass[prl,twocolumn,superscriptaddress]{revtex4}

\usepackage{graphicx}

\begin{document}
\newtheorem{theorem}{Theorem}
\newtheorem{acknowledgement}[theorem]{Acknowledgement}
\newtheorem{algorithm}[theorem]{Algorithm}
\newtheorem{axiom}[theorem]{Axiom}
\newtheorem{claim}[theorem]{Claim}
\newtheorem{conclusion}[theorem]{Conclusion}
\newtheorem{condition}[theorem]{Condition}
\newtheorem{conjecture}[theorem]{Conjecture}
\newtheorem{corollary}[theorem]{Corollary}
\newtheorem{criterion}[theorem]{Criterion}
\newtheorem{definition}[theorem]{Definition}
\newtheorem{example}[theorem]{Example}
\newtheorem{exercise}[theorem]{Exercise}
\newtheorem{lemma}[theorem]{Lemma}
\newtheorem{notation}[theorem]{Notation}
\newtheorem{problem}[theorem]{Problem}
\newtheorem{proposition}[theorem]{Proposition}
\newtheorem{remark}[theorem]{Remark}
\newtheorem{solution}[theorem]{Solution}
\newtheorem{summary}[theorem]{Summary}    
\def\r{{\bf{r}}}
\def\i{{\bf{i}}}
\def\j{{\bf{j}}}
\def\m{{\bf{m}}}
\def\k{{\bf{k}}}
\def\kt{{\tilde{\k}}}
\def\mt{{\hat{t}}}
\def\mG{{\hat{G}}}
\def\mg{{\hat{g}}}
\def\mGa{{\hat{\Gamma}}}
\def\mS{{\hat{\Sigma}}}
\def\mT{{\hat{T}}}
\def\K{{\bf{K}}}
\def\P{{\bf{P}}}
\def\q{{\bf{q}}}
\def\Q{{\bf{Q}}}
\def\p{{\bf{p}}}
\def\x{{\bf{x}}}
\def\X{{\bf{X}}}
\def\Y{{\bf{Y}}}
\def\F{{\bf{F}}}
\def\G{{\bf{G}}}
\def\bG{{\bar{G}}}
\def\mbG{{\hat{\bar{G}}}}
\def\M{{\bf{M}}}
\def\V{\cal V}
\def\tchi{\tilde{\chi}}
\def\tx{\tilde{\bf{x}}}
\def\tk{\tilde{\bf{k}}}
\def\tK{\tilde{\bf{K}}}
\def\tq{\tilde{\bf{q}}}
\def\tQ{\tilde{\bf{Q}}}
\def\si{\sigma}
\def\ep{\epsilon}
\def\hep{{\hat{\epsilon}}}
\def\al{\alpha}
\def\be{\beta}
\def\ep{\epsilon}
\def\bep{\bar{\epsilon}_\K}
\def\mep{\hat{\epsilon}}
\def\up{\uparrow}
\def\de{\delta}
\def\De{\Delta}
\def\up{\uparrow}
\def\dwn{\downarrow}
\def\ksi{\xi}
\def\etha{\eta}
\def\product{\prod}
\def\goto{\rightarrow}
\def\switch{\leftrightarrow}

\title{Zinc Impurities in the 2D Hubbard model}
\author{
Th.\ A.\ Maier
and
M.\ Jarrell
}
\address{University of Cincinnati, Cincinnati OH 45221, USA}

\begin{abstract}
We study the two-dimensional Hubbard model with nonmagnetic Zn impurities 
modeled by binary diagonal disorder using Quantum Monte Carlo within the 
Dynamical Cluster Approximation. With increasing Zn content we find a 
strong suppression of $d$-wave superconductivity concomitant to a reduction 
of antiferromagnetic spin fluctuations. $T_c$ vanishes linearly with Zn
impurity concentration. The spin susceptibility changes from pseudogap to 
Curie-Weiss like behavior indicating the existence of free magnetic moments 
in the Zn doped system.  We interpret these results within the RVB picture.
\end{abstract}

\maketitle

\paragraph*{Introduction}

Chemical substitution in high-$T_c$ superconductors provides a powerful 
probe into the complex nature of both the superconducting and normal 
state of these materials. 
Experiments substituting different impurities for Cu show that 
nonmagnetic impurities (Zn, Al) are just as effective in suppressing 
superconductivity as magnetic dopants (Ni, Fe) \cite{Xiao}. Based on 
Anderson's theorem \cite{And} these results are taken as strong 
evidence for an unconventional pairing state described by an anisotropic 
order parameter with nodes on the Fermi surface. Indeed, recently the 
characteristic fourfold symmetry of the $d_{x^2-y^2}$-wave order parameter 
was observed in the spatial variation of the local density of states 
near the Zn impurity using scanning tunneling microscopy \cite{Pan}.   

Nuclear magnetic resonance experiments show that even a nonmagnetic 
impurity substituted for Cu induces an effective magnetic moment residing 
on the neighboring Cu sites \cite{NMR}. In addition, the bulk 
susceptibility in impurity substituted underdoped cuprate superconductors 
shows Curie-Weiss like behavior irrespective of the magnetic structure of 
the dopants \cite{Xiao}. This indicates the existence of free 
magnetic moments in the impurity doped ${\rm CuO}_2$ planes. The formation 
of these moments with substitution of nonmagnetic impurities is 
explained to arise from breaking singlets\cite{Xiao,Finkel} in an 
antiferromagnetically correlated host by removing Cu spins.     

In this letter we focus on the suppression of superconductivity by Zn 
impurities and the change of the bulk magnetic susceptibility to a 
Curie-Weiss like behavior in underdoped systems. Despite a wide variety 
of theoretical studies addressing these issues, a complete understanding of 
the effects of Zn doping has not been achieved. These approaches are mostly 
based on the description of impurities embedded in a BCS host \cite{BCS_Zn} 
or on generalizations of the Abrikosov-Gorkov equations for nonmagnetic 
impurities in unconventional superconductors using phenomenological pairing 
interactions \cite{AG_Zn}. 
In order to capture effects like moment formation by substitution of 
nonmagnetic impurities,  it seems necessary to describe correlations and 
the scattering from impurities on the same footing within a microscopic 
approach.

The most widely applied technique developed to describe disordered systems 
is the Coherent Potential Approximation (CPA) \cite{CPA}. 
The CPA shares the 
same microscopic definition \cite{DCA_Jarrell1} as its equivalent for 
correlated clean systems, the Dynamical Mean Field Approximation 
(DMFA) \cite{DMFA}. Both approaches 
map the lattice system 
onto an effective impurity problem embedded in a host that represents the 
remaining degrees of freedom. This single-site approximation 
neglects interference effects of the scattering off different impurity 
sites (crossing diagrams) and correlation effects become purely local. 
Therefore, it inhibits a transition to a state described by a nonlocal 
($d$-wave) order parameter and thus is not appropriate for our 
investigations here.

The Dynamical Cluster Approximation 
(DCA) \cite{DCA_Hettler1,DCA_Hettler2,DCA_Maier1,DCA_Jarrell1} systematically 
incorporates nonlocal corrections to 
the CPA/DMFA by mapping the lattice system onto an embedded periodic cluster 
of size $N_c$. For $N_c=1$ the DCA is equivalent to the CPA/DMFA and by 
increasing the cluster size $N_c$ the length-scale of possible dynamical 
correlations can be gradually increased while the DCA solution remains in 
the thermodynamic limit.  In the clean limit the DCA applied to the Hubbard 
model has been shown to describe the essential low energy physics of the 
cuprates \cite{DCA_Jarrell2,DCA_Maier2,DCA_Jarrell3}: It captures the 
antiferromagnetic phase near half filling and the transition to a 
superconducting phase with $d_{x^2-y^2}$-wave order parameter at finite doping. 
In the normal state it exhibits non-Fermi liquid behavior in form of a 
pseudogap in the density of states and a suppression of spin excitations at 
low temperatures in the underdoped regime. 

In this letter we study the two-dimensional (2D) Hubbard model including a potential scattering 
term according to the chemistry of Zn impurities using the DCA. We will show that potential scattering by Zn impurities  
strongly suppresses superconductivity as well as spin fluctuations and changes the magnetic susceptibility 
to a Curie-Weiss like temperature dependence in the underdoped region.

\paragraph*{Formalism}
The correlated electrons in a $\rm CuO_2$ plane doped with Zn 
impurities are described by the 2D Hubbard model with diagonal disorder 
\begin{equation}
\label{eq:Ham}
H=-t\sum_{<ij>,\sigma}c^\dagger_{i\sigma}c^{}_{j\sigma}+U\sum_in_
{i\uparrow}n_{i\downarrow}+\sum_{i\sigma}{\epsilon_in_{i\sigma}}\,,
\end{equation}
where we used standard notation. The disorder induced by the Zn sites occurs 
in the local orbital energies $\epsilon_i$ which are independent quenched 
random variables distributed according to some specified distribution  
$P(\epsilon_1,\dots ,\epsilon_N)=\product_{i=1}^N P(\epsilon_i)$ where $N$ is the number of lattice sites.
For a concentration $x$ of Zn impurities we use the binary alloy distribution
\begin{equation}
P(\epsilon_i)=x\delta(\epsilon_i+V/2)+(1-x)\delta(\epsilon_i-V/2)\,,
\end{equation}
where $V$ is the energy difference between the Cu and Zn 3$d$ orbitals. 
  For simplicity we 
use site-independent values for the hopping integral $t$ and the Coulomb 
repulsion $U$. 


The microscopic derivation of the DCA algorithm was discussed in detail for 
correlated systems in \cite{DCA_Hettler2,DCA_Jarrell3}  and for disordered 
systems in \cite{DCA_Jarrell1}.
The DCA has a simple physical interpretation for systems where the intersite 
correlations have only short spatial range.  The corresponding self-energy 
may then be calculated on a coarse grid of $N_c=L_c^D$ selected $\K$ points 
only, where $L_c$ is the linear dimension of the cluster of $\K$ points. 
Knowledge of the momentum dependence on a finer grid may be discarded to 
reduce the complexity of the problem. To this end the first Brillouin zone 
is divided into $N_c$ cells of size $(2\pi/L_c)^D$ around the cluster momenta 
$\K$. The propagators used to form the self-energy are coarse grained or 
averaged over the momenta $\K+\tk$ surrounding the cluster momentum $\K$
\begin{eqnarray}
\label{eq:cgG}
\bG(\K,i\omega_n)&=&\frac{N_c}{N}\sum_{\tk}
G(\K+\tk,i\omega_n)\nonumber\\
&=&\frac{N_c}{N}\sum_{\tk}\frac{1}{i\omega_n-\epsilon_
{\K+\tk} - \Sigma(\K,i\omega_n)}\,.
\end{eqnarray}
In Eq.(\ref{eq:cgG}) we make use of the fact that the approximation 
of the lattice self-energy $\Sigma(\k)$ by the cluster self-energy 
$\Sigma(\K)$ optimizes the free energy \cite{DCA_Hettler2,DCA_Maier1}. 
$\Sigma(\K)$ summarizes the single-particle effects of the interaction 
term (second term) and the disorder term (third term) of the Hamiltonian, 
Eq.(\ref{eq:Ham}). The dispersion $\epsilon_{\K+\tk}$ denotes the spectrum 
of the kinetic part (first term) of Eq.(\ref{eq:Ham}). In order to avoid 
overcounting of self-energy diagrams on the cluster, the cluster excluded 
propagator
\begin{equation}
{\cal G}(\K)^{-1}=\bG(\K)^{-1}+\Sigma(\K)
\end{equation}
is used as bare propagator for the cluster problem. To diagonalize the 
disorder part it is convenient to perform a Fourier transform to cluster 
real space ${\cal G}_{ij}=1/N_c\sum_\K e^{i\K\cdot(\X_i-\X_j)}{\cal G}(\K)$ 
where the disorder is diagonal. The inverse bare cluster propagator for a 
particular disorder configuration $\{\epsilon_i\}$ is then given by
\begin{equation}
[{\cal G}_{\epsilon_1,\dots ,\epsilon_N}^{-1}]_{ij}=[{\cal
G}^{-1}]_{ij}-\epsilon_{i}\delta_{ij}\,,
\end{equation}
and is used to initialize a QMC simulation to calculate the effects of the 
Coulomb interaction $U$. The QMC result for the cluster Green function 
$G^c_{\epsilon_1 ,\dots ,\epsilon_N,ij}=G^c_{ij}[U,{\cal
G}_{\epsilon_1,\dots ,\epsilon_N,ij}]$ thus depends on the particular 
disorder configuration $\{\epsilon_i\}$. The disorder averaged cluster 
Green function is then obtained from the individual results by\begin{equation}
\label{eq:daGc}
G^c_{ij}=\langle G^c_{\epsilon_1,\dots ,\epsilon_N,ij}\rangle\,,
\end{equation}
where the average $<\dots>=\int\product_{i=1}^N{\rm d}\epsilon_iP(\epsilon_i)(\dots)$  is to be taken for 
a system of $N_c$ sites.  Note that in principal QMC calculations for 
all possible disorder configurations $\{\epsilon_i\}$ have to be carried 
through. However, contributions from configurations with $m$ Zn impurities 
on the cluster ($m\le N_c$) are weighted by a factor of $x^m(1-x)^{(N_c-m)}$ to the 
integral Eq.(\ref{eq:daGc}). Therefore it seems reasonable for small 
concentrations $x\ll 1$ and cluster sizes $N_c$  to consider only those 
configurations with none or only a single Zn impurity on the cluster and 
to neglect configurations with more than one Zn impurity. Then the disorder 
averaged cluster Green function $G^c_{ij}$ is obtained as a weighted sum 
from the two configurations as
\begin{equation}
\label{eq:ava}
G^c_{ij}=xN_cG^c_{1,ij}+(1-xN_c)G^c_{0,ij}\,,
\end{equation}   
where $G^c_{1/0,ij}$ denotes the QMC result for the cluster Green function 
for the configuration with one or zero Zn impurities on the cluster, 
respectively. The prefactors follow with $(1-x)^{N_c}\approx 1-xN_c$, 
$xN_c(1-x)^{N_c-1}\approx xN_c$ for $x<<1$ in linear approximation in the 
impurity concentration $x$.
The disorder averaged cluster Green function $G^c_{ij}$ is then transformed 
back to cluster reciprocal space to calculate a new estimate of the cluster 
self-energy
\begin{equation}
\Sigma(\K)={\cal G}(\K)^{-1}-G^c(\K)^{-1}\,,
\end{equation}
which is then used to repeat the steps starting from Eq.(\ref{eq:cgG}) 
until convergence is reached.

\paragraph*{Results}
We study the superconducting instability and magnetic properties of 
the Hamiltonian Eq.(\ref{eq:Ham}) in the intermediate coupling regime 
at 5\% hole doping ($n=0.95$) for different Zn impurity concentrations $x$.
We perform calculations for $N_c=4$, the smallest cluster size that 
allows for a transition to a $d_{x^2-y^2}$-wave superconducting state 
while preserving the lattice translational and point group symmetries. 
We set the hopping integral 
$t=0.25 {\rm eV}$ and the Coulomb repulsion $U=W=2 {\rm eV}$, where 
$W=8t$ is the bandwidth of the non-interacting system. We choose 
$V=5 {\rm eV} > W+U$ to simulate the nonmagnetic closed shell ($d^{10}$) 
configuration of the impurity site.

\begin{figure}[h]
\includegraphics*[width=3.3in]{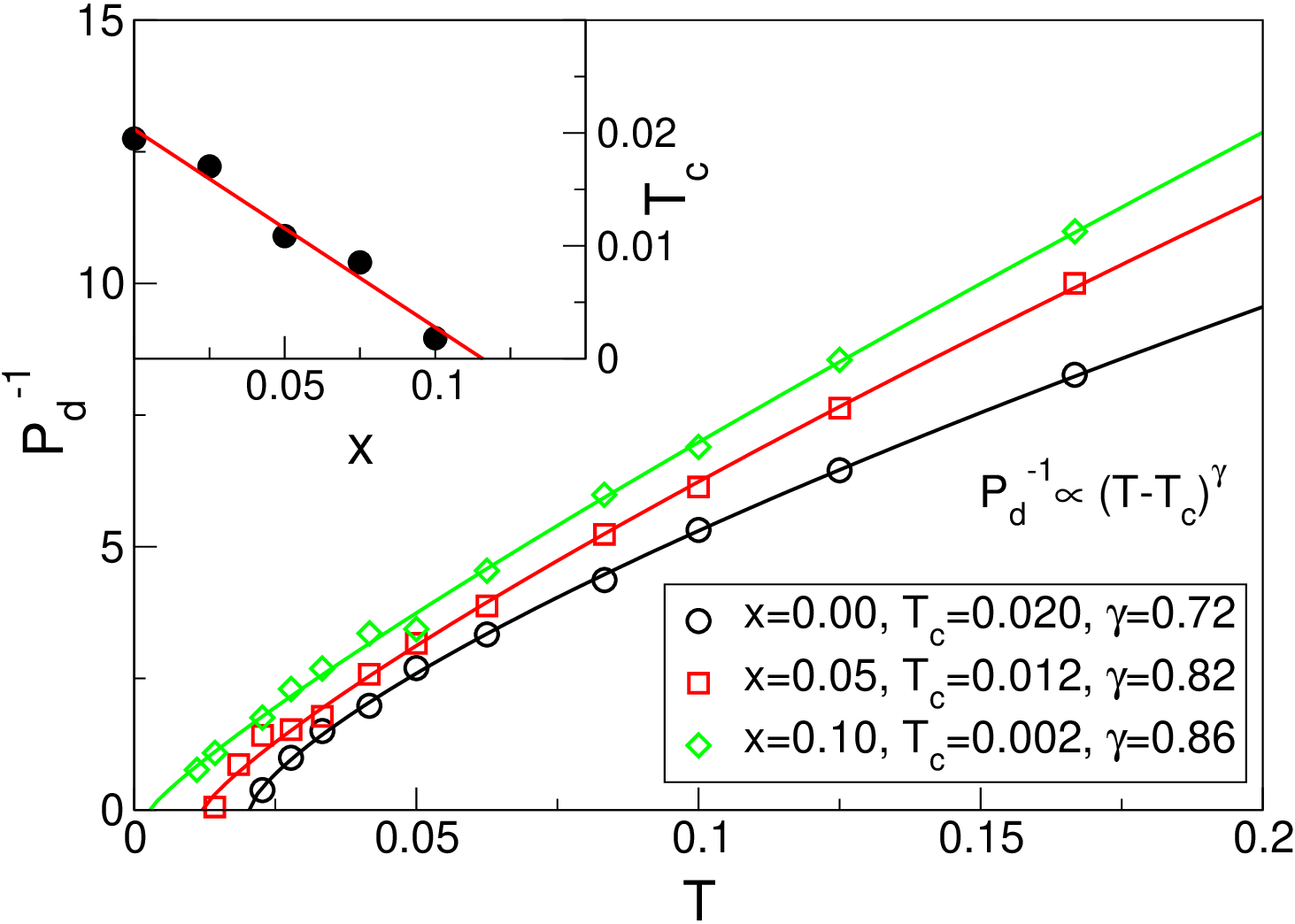}
\caption{The inverse pair-field susceptibility versus temperature when 
$N_c=4$, $U=W=2 {\rm eV}$, $V=5 {\rm eV}$ at 5\% doping for impurity 
concentrations $x=0$ (circles), $x=0.05$ (squares) and $x=0.10$ 
(diamonds). The solid lines represent fits to the function 
$P^{-1}_d\propto (T-T_c)^\gamma$. Inset: Critical temperature $T_c$ 
as a function of impurity concentration $x$.}
\label{fig:1}
\end{figure}

We searched for superconductivity with $s$-wave, extended $s$-wave, $p$- 
and $d$-wave order parameters. As we found for the clean system 
($x$=0) \cite{DCA_Jarrell2,DCA_Jarrell3}, only the $d_{x^2-y^2}$-wave 
pair field susceptibility for zero center of mass momentum diverges at 
low temperatures. Fig.\ref{fig:1} illustrates the  inverse pair field 
susceptibility $P^{-1}_d$ as a function of temperature $T$ for the impurity 
concentrations $x=0$ (circles), $x=0.05$ (squares) and $x=0.10$ (diamonds). 
The corresponding critical temperature $T_c$ is then calculated by 
extrapolating $P^{-1}_d(T)$ to zero using the function 
$P^{-1}_d\propto (T-T_c)^{\gamma}$ (represented by the solid lines 
in Fig.\ref{fig:1}).  $T_c$ is rapidly suppressed, and the critical 
exponent $\gamma$ increases, with increasing Zn content $x$. 
Consistent with experiments \cite{Xiao} this falloff of $T_c$ is linear 
in $x$ resulting in a critical impurity concentration $x_c \approx 0.10$, 
beyond which the instability to the superconducting 
phase disappears. Compared to experiments ($x_c\lesssim 0.05$) our 
calculation predicts a result about twice as high for the critical doping. 
However, considering the simplified description of the problem using 
a 2D Hubbard model with the complexity of the Zn impurities reduced 
to a model of diagonal disorder, we expect agreement only on a 
qualitative level.  

\begin{figure}[h]
\includegraphics*[width=3.3in]{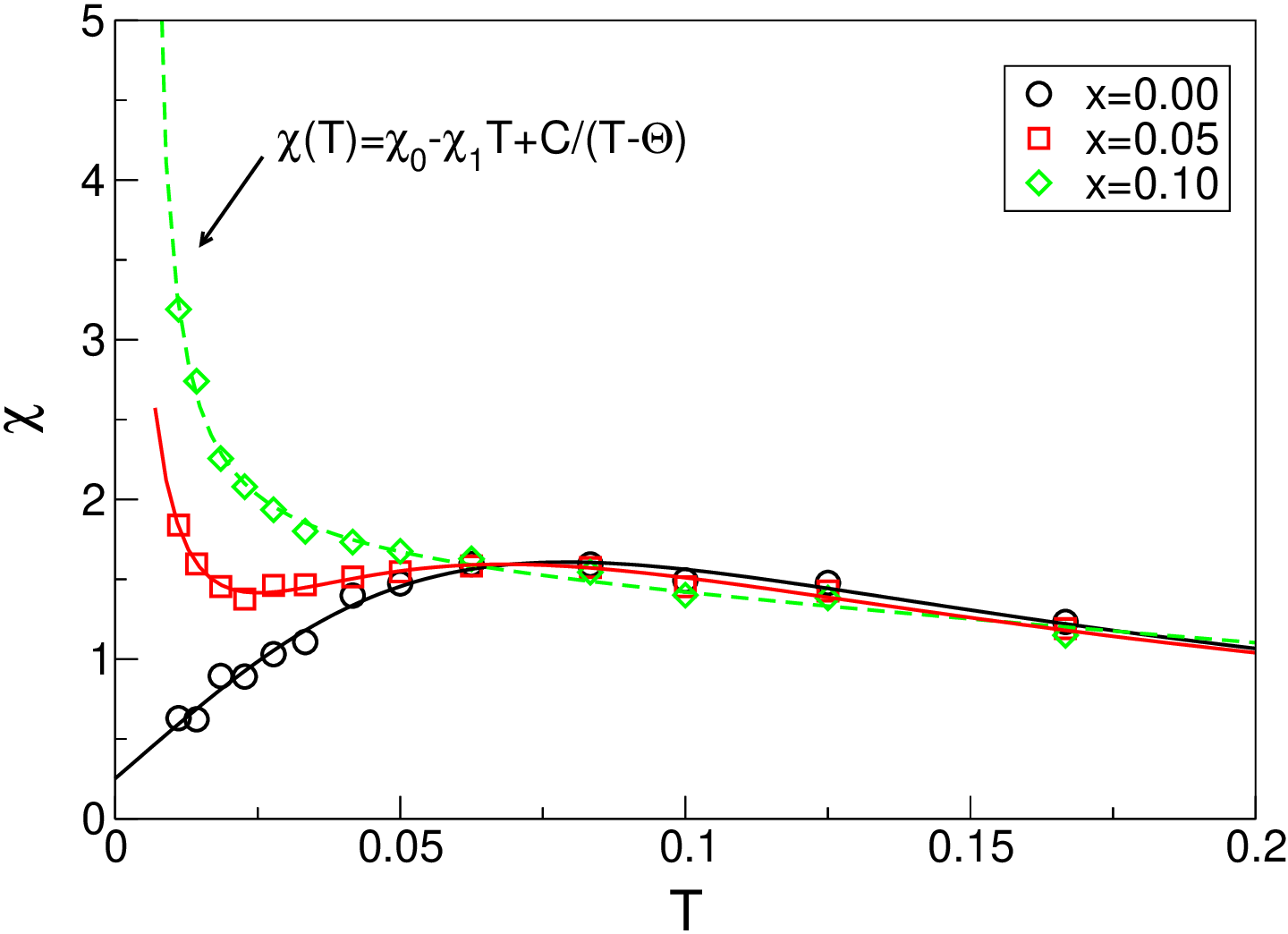}
\caption{The bulk magnetic susceptibility versus temperature when $N_c=4$, 
$U=W=2 {\rm eV}$, $V=5 {\rm eV}$ at 5\% doping for impurity concentrations 
$x=0$ (circles), $x=0.05$ (squares) and $x=0.10$ (diamonds). The solid 
lines are guides to the eye. The dashed line represents a fit to the 
function $\chi(T)=\chi_0-\chi_1T+C/(T-\Theta)$.}
\label{fig:2}
\end{figure}

In order to study the correlation between the suppression of 
superconductivity and possible moment formation by Zn substitution we 
study the bulk ($\q=(0,0)$) and the antiferromagnetic ($\q=(\pi,\pi)$) 
spin susceptibilities. Fig.\ref{fig:2} shows the bulk magnetic 
susceptibility $\chi(T)$ as a function of temperature for the same 
impurity concentrations $x$ used in Fig.\ref{fig:1}. 
As we demonstrated in \cite{DCA_Jarrell2}, the clean system exhibits an 
anomaly in $\chi(T)$ at a characteristic temperature $T^\star$, below 
which spin excitations are suppressed and $\chi(T)$ falls with decreasing 
temperature. $T^\star$ thus marks the onset of singlet formation due to 
strong short-ranged antiferromagnetic spin correlations. In agreement with 
experiments \cite{Xiao}, the substitution of impurities leads to a dramatic 
change of the low temperature behavior of the magnetic spin susceptibility. 
Despite the nonmagnetic nature of the impurity scattering term in our 
simulation, $\chi(T)$ displays an upturn at low temperatures as $x$ is 
increased. Moreover at $x=0.10$, $\chi(T)$ shows Curie-Weiss like 
behavior; i.e. it can be fit by the function 
$\chi(T)=\chi_0-\chi_1T+C/(T-\Theta)$, with $\Theta\approx 0$. This clearly 
indicates the formation of free magnetic moments due to the breaking of 
singlet bonds by the substitution of nonmagnetic impurities. 

\begin{figure}[h]
\includegraphics*[width=3.3in]{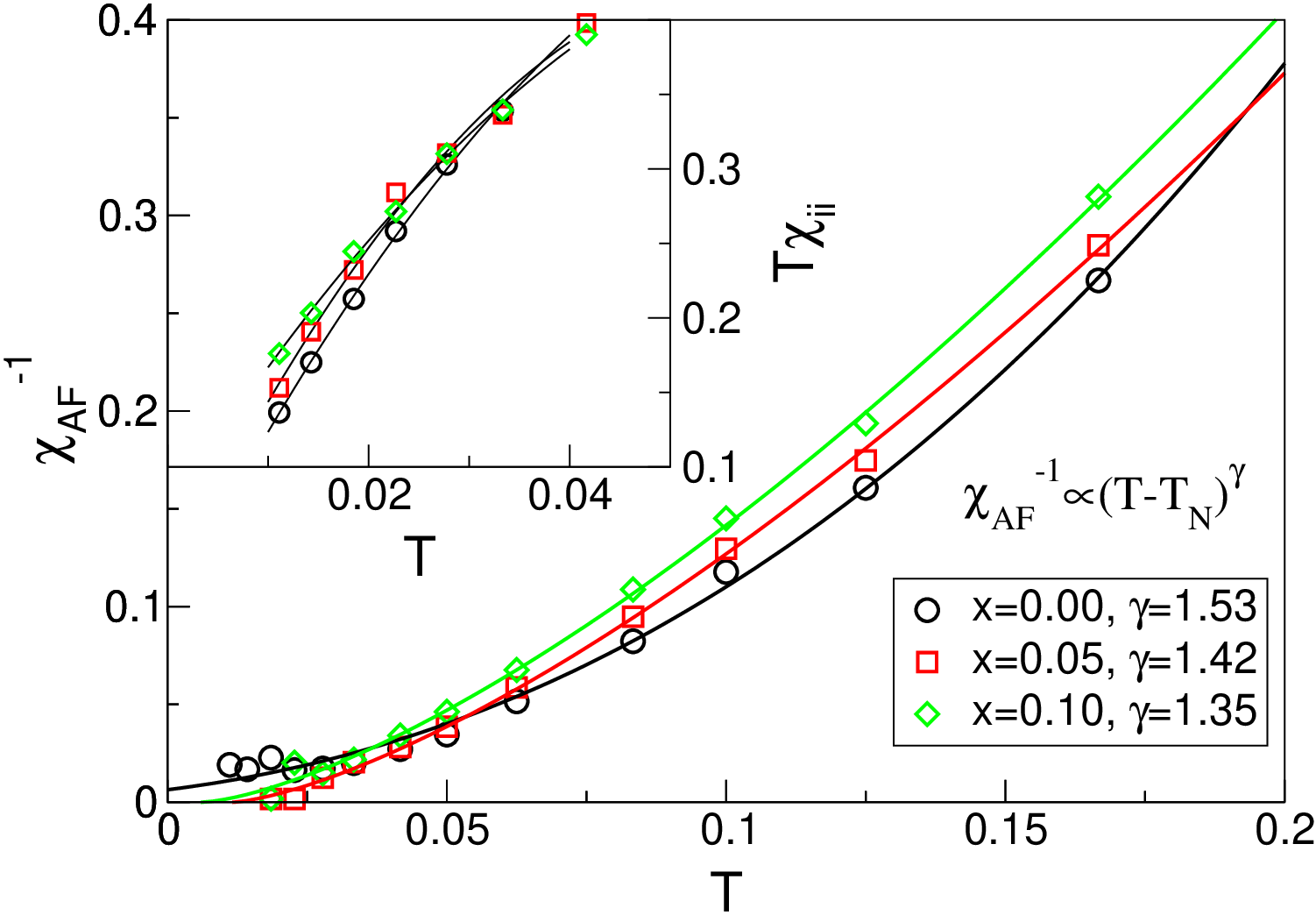}
\caption{The inverse antiferromagnetic ($\q=(\pi,\pi)$) spin susceptibility 
versus temperature when $N_c=4$, $U=W=2 {\rm eV}$, $V=5 {\rm eV}$ at 5\% 
doping for impurity concentrations $x=0$ (circles), $x=0.05$ (squares) 
and $x=0.10$ (diamonds). The solid lines represent fits to the function 
$\chi_{AF}^{-1}\propto (T-T_{\rm N})^\gamma$. Inset: The site and time 
averaged magnetic moment versus temperature; spin vacancies reduce 
fluctuations.  The solid lines are guides to the eye.}
\label{fig:3}
\end{figure}

The inverse antiferromagnetic spin susceptibility $\chi_{AF}(T)$ for the 
same parameter set as in Fig.\ref{fig:2} is illustrated in Fig.\ref{fig:3}. 
For the clean system (x=0) 5\% hole doping marks the critical doping, beyond 
which no transition to an antiferromagnetic state can be 
found \cite{DCA_Jarrell2}. As can be seen from the extrapolation of the 
circles in Fig.\ref{fig:3} the corresponding Ne\'{e}l temperature 
$T_{\rm N}=0$. As the concentration of impurities is increased we notice 
that for $T\gtrsim 0.05$  $\chi_{AF}$ decreases with $x$. 
However at lower temperatures ($T\lesssim 0.05$) $\chi^{-1}_{AF}$ 
for finite impurity concentration $x$ falls below the curve of the clean 
system and eventually goes through zero indicating a transition to the 
antiferromagnetic state.  Therefore it can be inferred that a finite 
concentration of impurities, i.e. spin vacancies, acts to enhance 
antiferromagnetic spin correlations.   
Our results for the site and time averaged local magnetic moment $T\chi_{ii}$ 
shown in the inset to Fig.\ref{fig:3} support this conjecture: At low 
temperatures this quantity increases with impurity concentration $x$ 
indicating that local moments get stabilized. It is important to note that 
recent NMR measurements on YBCO \cite{Jul} show a similar enhancement of 
antiferromagnetic spin correlations around Zn impurities at low temperatures. 

Similar results were also obtained in \cite{Bulut,Dagotto} where an 
antiferromagnetic Heisenberg model with a spin vacancy was studied. 
In the latter report, the enhancement of antiferromagnetic spin correlations 
was ascribed to {\em pruning} of singlets by the nonmagnetic impurity in an 
resonating-valence-bond picture (RVB). In the RVB picture, the short-ranged
order in the doped model is described as a collection of nearest-neighbor
singlet bonds which fluctuate as a function of time and space.  When a Zn 
impurity is introduced into the system, the singlet fluctuations on adjacent 
sites are suppressed since these sites now have one less neighbor to form 
short-ranged bonds with.  The suppression of singlet fluctuations on the 
sites adjacent to the Zn dopant also enhances moment formation on these sites.

We find further evidence of the suppression of spin fluctuations by Zn 
impurities in the decrease of the antiferromagnetic exponent $\gamma$ 
towards the mean-field result of one with increasing impurity content 
$x$. Concomitant with this behavior the corresponding exponent $\gamma$ 
for the pair-field susceptibility (cf.\ Fig.\ref{fig:1}) increases towards 
one. From this observation we conclude that the suppression of 
superconductivity in Zn doped systems originates in the suppression of 
antiferromagnetic spin fluctuations that mediate pairing.       
     
\paragraph*{Summary}
In this letter we have used the 2D Hubbard model with binary diagonal disorder 
to study the effects of nonmagnetic (Zn) impurities on high-temperature 
superconductivity. We find that Zn impurities strongly suppress 
superconductivity. As evidenced by the increasing mean-field character of 
the pairfield and antiferromagnetic susceptibilities with Zn substitution, 
spin fluctuations that mediate pairing get suppressed. Consistent with 
experiments, $T_c$ decreases linearly with impurity concentration. With 
increasing Zn content we further find a change of the magnetic susceptibility 
to Curie-Weiss like behavior indicative of the existence of free magnetic 
moments.  

Our results can be understood within the RVB picture. The nonmagnetic 
impurities break local singlets and thus generate unpaired spins.
A concomitant suppression of antiferromagnetic spin fluctuations that 
mediate pairing results from the pruning of RVB states \cite{Dagotto}. 
Consequently, superconductivity gets strongly suppressed with increasing 
Zn content.


\paragraph*{Acknowledgements} We acknowledge useful conversations with
W.\ Putikka,
D.J.\ Scalapino,
Y.\ Wang and M. Vojta.
This work was supported by NSF grant DMR-0073308.
This research was supported in part by NSF 
cooperative agreement ACI-9619020 through computing resources provided 
by the National Partnership for Advanced Computational Infrastructure 
at the Pittsburgh Supercomputer Center.


\end{document}